\documentclass{acm_proc_article-sp}[10pt]

\newif\ifpdf
\ifx\pdfoutput\undefined
   \pdffalse     
\else
   \pdfoutput=1  
   \pdftrue
\fi

\usepackage{times}

\ifpdf
\usepackage[pdftex]{graphicx}
\else
\usepackage{graphicx}
\fi

\usepackage{xspace}

\ifpdf
\usepackage[pdftex,bookmarks=false,
            plainpages=false,naturalnames=true,
            colorlinks=true,pdfstartview=FitV,
            linkcolor=blue,citecolor=blue,urlcolor=blue]{hyperref}
\else
\usepackage[dvips]{hyperref}
\fi

\newcommand{\myhref}[2]{\ifpdf\href{#1}{#2}\else\htmladdnormallinkfoot{#2}{#1}\fi}

\newcommand{\obj}{\footnotesize \begin{alltt}}
\newcommand{\jbo}{\end{alltt} \normalsize}

\newcommand{\isa}{\ensuremath{<}}

\newcommand{\hasa}{\ensuremath{\supset}}
\newcommand{\compose}{\ensuremath{\circ}}
\newcommand{\prcompose}{\ensuremath{\otimes}}
\newcommand{\cocompose}{\ensuremath{\oplus}}
\newcommand{\partinterp}{\ensuremath{\hookrightarrow}}

\newcommand{\fullinterp}{\ensuremath{\leadsto}}

\newcommand{\realizes}{\ensuremath{:}}
\newcommand{\kindnamefont}[1]{\textsc{#1}}
\newcommand{\knf}[1]{\kindnamefont{#1}}

\newcommand{\parent}{\ensuremath{\isa_{p}}}
\newcommand{\partof}{\ensuremath{\subset_{p}}}

\newcommand{\fullequiv}{\ensuremath{\eqcirc}}
\newcommand{\partequiv}{\ensuremath{\lessdot}}

\newcommand{\isakindof}{\ensuremath{\isa_{r}}}

\newcommand{\kLoop}{\knf{Loop}}

\begin{document}

\conferenceinfo{FSE-10}{2002 Charleston, South Carolina, USA}
\CopyrightYear{2002}

\title{Semantic Properties for Lightweight Specification in \\
  Knowledgeable Development Environments}
\numberofauthors{1}
\author{
\alignauthor Joseph R.~Kiniry\\
       \affaddr{Department of Computer Science}\\
       \affaddr{California Institute of Technology}\\
       \affaddr{Mailstop 256-80}\\
       \affaddr{Pasadena, CA 91125}\\
       \email{kiniry@acm.org}
}

\maketitle

\begin{abstract}
  
  Semantic properties are domain-specific specification constructs used to
  augment an existing language with richer semantics.  These properties are
  taken advantage of in system analysis, design, implementation, testing,
  and maintenance through the use of documentation and source-code
  transformation tools.  Semantic properties are themselves specified at
  two levels: loosely with precise natural language, and formally within
  the problem domain.  The refinement relationships between these
  specification levels, as well as between a semantic property's use and
  its realization in program code via tools, is specified with a new formal
  method for reuse called kind theory.
  
\end{abstract}


                [design languages]
                [preprocessors]
                

\vspace{1in}

\section{Introduction}
\label{sec:introduction}

Ad hoc constructs and local conventions have been used to annotate program
code since the invention of programming languages.  The purpose of these
annotations is to convey extra programmer knowledge to other system
developers and future maintainers.  These comments usually fall into that
gray region between completely unstructured natural language and formal
specification.

Invariably, such program comments rapidly exhibit ``bit rot''.  Over time,
these comments, unless well maintained by documentation specialists,
rigorous process, or other extra-mile development efforts, become
out-of-date.  They are the focus for the common mantra: an incorrect
comment is worse than no comment at all.

Recently, with the adoption and popularization of light\-weight
documentation tools in the literate programming tradition~\cite{Knuth92,
  KnuthLevy93}, an ecology of semi-structured comments is flourishing.  The
rapid adoption and popularity of Java primed interest in semi-structured
comment use via the Javadoc tool.  Other similar code-to-documentation
transformation tools have since followed in volume including Jakarta's
Alexandria, Doxygen, and Apple's HeaderDoc.
\myhref{http://www.sf.net/}{Source\-Forge} reports thirty-six projects with
``Javadoc'' in the project summary.
\myhref{http://www.freshmeat.net/}{Fresh\-Meat} reports another
thirty-five, with some overlap.

While most of these systems are significantly more simple than Knuth's
original CWEB, they share two key features.  

First, they are easy to learn, since they necessitate only a small change
in \emph{convention} and \emph{process}.  Rather than forcing the
programmer to learn a new language, complex tool, or imposing some other
significant barrier to use, these tools actually \emph{reward} the
programmer for documenting her code.  

Second, a culture of documentation is engendered.  Prompted by the example
of vendors like Sun, programmers enjoy the creation and use of the
attractive automatically-generated documentation in a web page format.
This documentation-centric style is only strengthened by the exhibitionist
nature of the Web.  Having the most complete documentation is now a point
of pride in some Open Source projects; a state of affairs we would not have
guessed at a decade ago.

The primary problem with these systems, and the documentation and code
written using them, is that \emph{even semi-structured comments have no
  semantics}.  Programmers are attempting to state (sometimes quite
complex) knowledge but are not given the language and tools with which to
communicate this knowledge.  And since the vast majority of developers are
unwilling to learn a new, especially formal, language with which to convey
such information, we must look for a happy-medium of \emph{informal
  formality}.

That compromise, the delicate balance between informality and formality,
and the lightest-weight aspect of our \emph{Knowledgeable Software
  Engineering} program, is what we call \emph{semantic properties}.

Semantic properties are domain-independent documentation constructs with
intuitive formal semantics that are mapped into the semantic domain of
their application.  Semantic properties are used as if they were normal
semi-structured documentation. But, rather than being ignored by compilers
and development environments as comments typically are, they have the
attention of augmented versions of such tools.  Semantic properties embed
a tremendous amount of concise information wherever they are used without
imposing the often insurmountable overhead seen in the introduction of new
languages and formalisms for similar purposes.

\section{Semantic Properties}
\label{sec:semantic-properties}

The original inspiration for semantic properties came from three sources:
the use of tags, (e.g., \texttt{@author} and \texttt{@param}), in the
Javadoc system, the use of annotations and pragmas in languages like Java
and C for code transformation and guided compilation, and indexing clauses
in Eiffel.  All of these systems have a built-in property-value mechanism,
one at the documentation level and one within the language syntax itself,
that is used to specify semi-structured information.

In Java, tags are the basic realization of our semantic properties.  They
are used for documentation and formal specification, as we will discuss in
more detail in Section~\ref{sec:java}.  Tags are not part of the language
specification.  In fact, they are entirely ignored by all Java compilers.

Annotations and pragmas come in the form of formal tags used for some
Design by Contract tools like
\myhref{http://semantik.informatik.uni-oldenburg.de/~jass/}{Jass} which
happen to be realized in Eiffel with first-class keywords like
\emph{require}, \emph{ensure}, and \emph{invariant}.

Eiffel provides first-class support for properties via \emph{indexing
  clauses}.  An Eiffel file can contain arbitrary property-value pairs
inside of \texttt{indexing} blocks.  This information is used by a variety
of tools for source code search, organization, and documentation.

\vspace{0.5in}

\subsection{Documentation Semantics}
\label{sec:docum-semant}

Recently, Sun has started to extend the semantics of these basic properties
with respect to language semantics, particularly with regards to
inheritance.  If a class $C$ inherits from a parent class $P$, and $P$'s
method $m$ has some documentation, but $C$'s overridden or effective (in
the case where $P$ and/or $m$ is \emph{abstract}) version of $m$ does not,
then Javadoc \emph{inherits} $C.m$'s documentation for $P.m$, generating
the appropriate comments in Javadoc's output.

This change in behavior of the tools is an implicit change in the semantics
of the documentation.  While straightforward and useful in this example,
the meaning of such inheritance is undocumented and often unclear.

The situation in Eiffel is less confusing.  The semantics of properties, as
realized by indexing clauses and formal program annotation via contracts,
are defined in the language standard~\cite{Meyer90-ETL}.  

Even so, no mechanism exists in either system for extending these
semi-structured comments with new semantics beyond simple plug-ins for
documentation (e.g., doclets in Java and translators in EiffelStudio).

Also, the semantics of current annotations are entirely specified within a
particular language or formalism.  No general purpose formalism has been
used to express their extra-model semantics.

\subsection{Semantics of Semantic Properties}
\label{sec:semant-semant-prop}

We specify the semantics of semantic properties in a new formalism called
\emph{kind theory}.  Kind theory is a logic used to describe, reason about,
and discover reusable assets of arbitrary sorts.  Kind theory is an
higher-order, autoepistemic\footnote{Auto-epistemic: ``representing or
  discussing self-knowledge''.}, paraconsistent\footnote{Paraconsistent:
  ``explicitly representing and reasoning with potential and transitory
  logic inconsistency''.}, categorical logic with a type theoretic and
algebraic model, and is described in full detail in Kiniry's
dissertation~\cite{Kiniry02-PhD}.

\subsubsection{A Brief Overview of Kind Theory}
\label{sec:brief-overview-kind}

\emph{Kind} are classifiers used to describe reusable assets like program
code, components, documentation, specifications, etc. \emph{Instances} are
realizations of kind---actual embodiments of these classifiers.  For
example, the paperback ``The Portrait of the Artist as a Young Man'' by
James Joyce is an \emph{instance} of \emph{kinds} $\knf{PaperbackBook}$,
$\knf{EnglishDocument}$, and others.

In the context of semantic properties, our kinds are the semantic
properties as well as the programming language constructs to which the
properties are bound.  Our instances are the specific realizations of these
kinds within a particular input structure, typically a programming or
specification language.

Kinds are described structurally using our logic in a number of ways.
Classification is covered with the inheritance operators $\isa$ and
$\parent$; structural relationships are formalized using the inclusion
operators $\partof$ and $\hasa$, equivalence has several forms $\fullequiv$
and $\partequiv$; realization, the relationship between instances and kind,
is formalized with the operators $\isakindof$ and $\realizes$; composition
is captured in several forms, $\prcompose$, $\cocompose$, and $\compose$;
and interpretation, the translation of kind to kind or instances to
instances, is realized with the operators $\partinterp$ and $\fullinterp$.

Semantics are specified in an autoepistemic fashion using what are called
\emph{truth structures}.  Truth structures come in two forms:
\emph{claims} and \emph{beliefs}.

\emph{Claims} are stronger than beliefs.  A mathematically proven statement
that is widely accepted is a claim.  This phrasing is used because, for
example, there are theorems that have a preliminary proof but are not yet
widely recognized as being true.

A statement that is universally accepted, but not necessarily
mathematically proven, is also a claim.  Claims are not necessarily
mathematical formulas.  The statement ``the sun will rise tomorrow'' is
considered by the vast majority of listeners a true and valid statement,
thus is classified as a claim rather than as a belief.

\emph{Beliefs}, on the other hand, range in surety from \emph{completely
  unsure} to \emph{absolutely convinced}.  No specific metric is defined
for the degree of conviction, the only requirement placed on the associated
belief logic is that the belief degree form a partial order.

We use kind theory to specify semantic properties because it provides us
with an excellent model-independent (i.e., it is not bound to some specific
programming language) reuse-centric formalism.  Kind theory's whole
purpose is the specification of such reusable concepts.

We have insufficient space to summarize kind theory here, so we will simply
provide some basic definitions, examples, and motivation of its use within
the context of semantic properties.

\subsubsection{Kind Theory and Semantic Properties}
\label{sec:kind-theory-semantic}

Using kind theory, we specify the semantic relationships between
specifications and their realization.  With regards to semantic properties,
these relation\-ships formally explain which properties exist, how they can
be structured, how they can be applied in a specific language or system,
and how they are interpreted into alternative forms like documentation and
test code.

First, we denote the classification relationships between kinds using
inheritance operators.  Properties are classified using standard conceptual
data modeling~\cite{HerzigGogolla92, Wille97} and onto\-logy
en\-gin\-eer\-ing~\cite{Bench-CaponMalcolm99, Gruber00, VandervetMars98}
techniques into a kind hierarchy.  Details about these classifications for
semantic properties are discussed in Section~\ref{sec:prop-class}.

Next, the structural relationships between kinds are specified using the
inclusion operators.  Structural relationship denote the contexts in which
a kind can be used and how kinds are composed to create new kinds.  We
discuss structural relations in more detail in Section~\ref{sec:context}.

Finally, equivalence relations and interpretations are defined on kinds.
Equivalence relations help refine concepts embodied as kinds for particular
domain models by folding and simplifying kind hierarchies.  They also let
the user pick representative structures, called \emph{canonical forms},
that are used to represent and reason about (semi-)equivalent kind.

Interpretations, which are structure-preserving functions, are defined to
help capture notions of inheritance, equivalence, and other inter-domain
translations.

The formal aspects of kind theory, the specification of kind domains for
things like semantic properties, is performed by an expert.  The typical
software engineer never needs to learn or witness the formalism to benefit
from its availability.

\paragraph{A Kind Example}
Consider a loop in any standard programming language.  Loops come in many
syntactic forms.  For example, in the C programming language there are
three primary loop constructs: \texttt{for}, \texttt{while}, and
\texttt{do}.  Fundamentally, all loop constructs are equivalent to each
other at some abstraction level; they are each just syntactic variations on
a common theme.  That theme is specified by the kind $\kLoop$.  

We classify loops as a computational structure.  This classification states
that the notion of a loop is going to be bound to a specific syntax and can
be interpreted in some programmatic context.  We state this relation as
               $$\kLoop \isa \knf{ComputationalStructure}$$
By the rules of inheritance, all the structure inherent in the
parent kind, that of $\knf{ComputationalStructure}$, is realized in the
child kind $\kLoop$ as well.

Additionally, an interpretation exists of the form
            $$\kLoop \partinterp \knf{ComputationalStructure}$$
that takes kinds or instances of loops to kinds or instances of
computational structures, respectively.  This function is called a
\emph{partial interpretation} because it eliminates all of the semantics of
loops that differentiates them from general computational structures.
Interpretations are realized by categorical forgetful functors in kind
theory.

The structure of loops is straightforward.  Each loop has an initial state,
an increment function, a guard, and a body.  We state this kind
theoretically as
\begin{gather*}
  \knf{InitialState} \partof \kLoop \\
  \knf{IncrementFunction} \partof \kLoop \\
  \knf{GuardPredicate} \partof \kLoop \\
  \knf{LoopBody} \partof \kLoop
\end{gather*}
Each of these kind, in turn, has its own structure associated with it.
$\knf{GuardPredicate} \isa \knf{Predicate}$, for example.

Interpretations let us do two primary things.  First, we use
interpretations to translate among different forms of loops, converting a
\texttt{while} loop into a \texttt{do} loop, for example.  Second, they are
used for interpreting the generic semantics of loops in a specific language
or formal context.

For instance, a generic specification of a loop instance can be translated
to and from a specific syntactic structure realized in the Java programming
language.  Additionally, a formally specified loop, (the kind
$\knf{FormalLoop}$), complete with a loop variant function and invariant
predicate, can be translated into a proof structure in our logical
framework.  This opens up the opportunity for statically proving the
correctness of such formally specified loops.

All of the above details are fully formalized in a \emph{kind domain} in
the basic kind system~\cite{Kiniry02-PhD}.

A \emph{kind domain} is simply a set of kind and instances that are
specific to some domain of knowledge.  In this case, that domain is one of
language-generic computational structures, and we call that domain
$\knf{ComputationalStructures}$.

A \emph{kind system} is an actual computational system that realizes kind
theory.  Our first realization is witnessed both in a logical framework,
that of SRI's Maude~\cite{ClavelDuranEtal99-Metatool}, as well as in
software engineering tools like the \myhref{http://www.jiki.org/}{Jiki}
(used as an open, collaborative knowledge repository), the
JPP~\cite{KiniryCheong98}, and the \myhref{http://ebon.sf.net/}{EBON tool
  suite} (a design model checker).  Some of these tools are discussed in
more detail in
Sections~\ref{sec:tool-support}~and~\ref{sec:embedd-semant-prop}.

\subsection{Properties and Their Classification}
\label{sec:prop-class}

We have defined elsewhere\footnote{The original specification of these
  properties was in the Infospheres Java Coding Standard
  (\href{http://www.infospheres.caltech.edu/})
  {http://www.infospheres.caltech.edu/}.  That standard has since been
  refined and broadened.  More recent versions are available at
  KindSoftware (\href{http://www.kindsoftware.com/}
  {http://www.kindsoftware.com/}).}  thirty-five semantic properties.  All
semantic properties are enumerated in
Table~\ref{tab:The_Full_Set_of_Semantic_Properties} in the appendix.  Since
we have only a limited amount of space in this paper, we will summarize
some of the more interesting properties, their semantics, and our
experiences with their use in a number of software engineering projects,
large and small, over the last five years.

To derive our core set of semantic properties, we abstracted and unified
the existing realizations that we have used in two languages for many
years.  First, we gathered the set of predefined Javadoc tags, the standard
Eiffel indexing clauses, and the set of basic formal specification
constructs.  After that set was made self-consistent, that is, duplicates
were removed, semantics were weakened across domains for the
generalization, etc., we declared it the \emph{core set} of semantic
property kind.

These properties were then classified according to their general use and
intent.  The classifications are: \emph{meta-information}, \emph{pending
  work}, \emph{contracts}, \emph{concurrency}, \emph{usage information},
\emph{versioning}, \emph{inheritance}, \emph{documentation},
\emph{dependencies}, and \emph{miscellaneous}.  This classification is
represented using kind theory's inheritance operators.


Many of these semantic properties are used solely for documentation
purposes.  The \texttt{title} property documents the title of the project
with which a file is associated; the \texttt{description} property provides
a brief summary of the contents of a file.  We call these \emph{informal}
semantic properties.

Another set of properties are used for specifying non-program\-matic
semantics.  By ``non-programmatic'' we mean that the properties have
semantics, but they are not, or cannot, be expressed in program code.  For
example, labeling a construct with a \texttt{copyright} or \texttt{license}
property specifies some legal semantics.  Tagging a method with a
\texttt{bug} property specifies that the method has some erroneous behavior
that is described in detail in an associated bug report.  We call these
properties \emph{semi-formal} because they have a semantics, but outside of
the domain of software.

Finally, the balance of the properties specify structure that is
programmatically testable, checkable, or verifiable.  Basic examples of
such properties are \texttt{require} and \texttt{ensure} tags for
preconditions and postconditions, \texttt{modifies} tags for side-effect
semantics, and the \texttt{concurrency} and \texttt{generate} tags for
concurrency semantics.  These properties are called \emph{formal} because
they can be realized by a formal semantics.

The KindSoftware coding standard~\cite{KS-Code_Standard01} summarizes our
current set of semantic properties.  Each property has a syntax, a correct
usage domain, and a natural language summary.  The formalization of
semantic properties is found in Kiniry's dissertation~\cite{Kiniry02-PhD}.

\subsection{Context}
\label{sec:context}

Each property has a legal scope of use, called its \emph{context}.
Contexts are defined in a coarse, language-independent fashion using
inclusion operators in kind theory.  Contexts are comprised of
\emph{files}, \emph{modules}, \emph{features}, and \emph{variables}.

\emph{Files} are exactly that: data files in which program code resides.
The scope of a file encompasses everything contained in that file.

A \emph{module} is some large-scale program unit.  Modules are typically
realized by an explicit module- or class-like structure.  Examples of
modules are classes in object-oriented systems, modules in languages of the
Modula and ML families, packages in the Ada lineage, etc.  Other words and
structures typically bound to modules include units, protocols, interfaces,
etc.

\emph{Features} are the entry point for computation.  Features are often
named, have parameters, and return values.  Functions and procedures in
structured languages are features, as are methods in object-oriented
languages, and functions in functional systems.

Finally, \emph{variables} are program variables, attributes, constants,
enumerations, etc.  Because few languages enforce any access principles for
variables, variables can vary in semantics considerably.

Each property listed in the appendix has a legal context.  The context
\emph{All} means that the property can be used at the file, module,
feature, or variable level.  Additional contexts can be defined, extending
the semantics of contexts for new programming language constructs that need
structured documentation with properties.

\subsection{Visibility}
\label{sec:visibility}

In languages that have a notion of \emph{visibility}, a property's
visibility is equivalent to the visibility of the context in which it is
used, augmented by domain-specific visibility options expressed in kind
theory.

Typical basic notions of visibility include \emph{public}, \emph{private},
\emph{children} (for systems with inheritance), and \emph{module} (e.g.,
Java's \emph{package} visibility).  More complex notions of visibility are
exhibited by C++'s notion of \emph{friend} and Eiffel's class-based feature
scoping.

Explicit visibilities for semantic properties are used to refine the notion
of specification visibility for organizational, social, and formal reasons.

For example, a subgroup of a large development team might choose to expose
some documentation for, and specification of, their work only to specific
other groups for testing, political, or legal reasons.

On the social front, new members of a team might not have yet learned
specific tools or formalisms used in semantic properties, so using
visibility to hide those properties will help avoid information overload.

Lastly, some formal specification, especially when viewed in conjunction
with standard test strategies (e.g., whitebox, greybox, blackbox, unit
testing, scenario-based testing), has distinct levels of visibility.  For
example, testing the postcondition of a private feature is only reasonable
and permissible if the testing agent is responsible for that private
feature.

\subsection{Inheritance}
\label{sec:inheritance}

Semantic properties also have a well-defined notion of \emph{property
  inheritance}.  Once again, we do not want to force new and complicated
extra-language semantics on the software engineer.  Therefore, property
inheritance semantics match those of the source language in which the
properties are used.  Our earlier discussion of basic comments for Java
methods (a feature property context) is an example of such property
inheritance.

These kinds of inheritance come in two basic forms: \emph{replacement} and
\emph{augmentation}.

The \emph{replacement} form of inheritance means that the parent property
is completely replaced by the child property.  An example of such semantics
are feature overriding in Java and the associated documentation semantics
thereof.

\emph{Augmentation}, on the other hand, means that the child's properties
are actually a \emph{composition} of all its parents' properties.  These
kinds of composition come in several forms.  The most familiar is the
standard substitution principle-based type semantics~\cite{LiskovWing93} in
many object-oriented systems, and the Hoare logic/Dijkstra calculus-based
semantics of contract refinement~\cite{Meyer92b}.

We can express these formal notions using kind theory because it is
embedded in a complete logical framework.  For example, we can
automatically reason about the legitimacy of specification refinement much
like Findler and Felleisen discuss in~\cite{FindlerFelleisen01}.

\subsection{Tool Support}
\label{sec:tool-support}

We have used these semantic properties for the last five years.  We have
found that, while an explicit adopted coding standard, positive feedback
via tools and peers, and course grade and monetary rewards goes a long way
toward raising the bar for documentation and specification quality, these
social aspects are simply not enough.  Process does help, regular code
reviews and pair programming in particular, but tool support is critical to
maintaining quality specification coverage, completeness, and consistency.

Templates were the first step taken.  We have used raw documentation and
code templates in programming environments ranging from \emph{vi} to
\emph{emacs} to \emph{jEdit}.  But templates only help prime the process,
they do not help maintain the content.

Code and comment completion also helps.  Completion is the ability of an
environment to use partial input to derive a typically more lengthy full
input.  We have experimented with augmented versions of completion in
\emph{emacs}, for example.

Likewise, documentation lint checkers, particularly those embedded in
development environments and documentation generators are also useful.  We
view source text highlighting, as in \emph{font-lock} mode in emacs, as an
extremely weak form of lint-checking.  The error reports issued by Javadoc
and its siblings are a stronger form of lint-checking and are quite useful
for documentation coverage analysis, especially when a part of the regular
build process.  Finally, scripts integrated into a revision control system
provide a ``quality firewall'' to a source code repository in much the same
fashion.

We believe that more can and should be done.  Our approach is to build and
use what we call \emph{Knowledgeable Development Environments} (KDEs).
These development environments use knowledge representation and formal
reasoning behind the scenes to help the user work smarter and not harder.

We have started work on such environment.  By extending powerful emacs
modes and tools that are part of our initial development environment (e.g.,
XEmacs coupled with the object-oriented browser, hyperbole, JDE, semantic,
and speedbar) with a kind system, we hope to raise the bar on development
environments.

\subsubsection{Current Work on KDEs}
\label{sec:current-work-kdes}

The first two features that we plan to implement are \emph{documentation
  inheritance} and \emph{perspective}.

Eiffel development environments contain tools that provide what are called
the \emph{flat}, \emph{short}, and \emph{contract} views of a class.  Flat
forms show the flattened version of a class---all inherited features are
flattened into a single viewpoint.  The short form eliminates the
implementation of all methods so that the developer can focus on a class's
interface.  The contract form is like the short form except the contracts
of the class (feature preconditions and postcondition, and class
invariants) are shown.  These forms can be combined, thus \emph{flat short}
or \emph{flat contract} forms have the obvious meanings.

\emph{Knowledgeable documentation inheritance} is an extended version of
such views.  Rather than manually program the semantics of the
``flattening'' operation, our formal specification in kind theory
automatically interprets the appropriate instances into a new form for
rendering within the knowledgeable development environment.  And because
such interpretations are often fully reversible, the flattened forms can be
edited and the changes will properly ``percolate'' to their original source
locations.

\emph{Perspectives} enable the user to specify which role(s) she is in
while interacting with the kind-enabled system.  Since kind theory is
autoepistemic, the specification of a role (represented by an \emph{agent}
within the theory) permits automatic ``filtering'' of information according
to, for example, visibility rules as discussed in
Section~\ref{sec:visibility}.  This user-centric filtering of information,
much like narrowing modes within Emacs, helps the user focus on the problem
at hand, ignoring all information that she either is not interested in,
concerned with, or should not see.

These are only two of our ideas for how to expose the user-centric aspects
of kind theory via development environments, incorporating the use of
semantic properties throughout.

\section{Embedding Semantic Properties}
\label{sec:embedd-semant-prop}

When a semantic property is bound to a particular instance, for example, an
\texttt{@author} tag is used in some Java source code, what does this
formally mean beyond questions of structural conformance?  How do these
semantic properties help guide the development process and exercise the
system during testing?  How do new tools take advantage of these
properties?

First, we have to \emph{embed} the semantic properties into the language in
which we are working.  Second, we need to define domain-specific semantics
using kind interpretations.  Lastly, we use kind theory's belief truth
structures to guide program development.

We will first look at syntactic embedding for two programming and one
specification language.  In the latter parts of this next section we will
address the other two points.

\subsection{Programming Languages}
\label{sec:progr-lang}

We have used semantic extensions in two programming languages: Java and
Eiffel.

\subsubsection{Java}
\label{sec:java}

Semantic properties are embedded in Java code using Javadoc-style comments.
This makes for a simple, parseable syntax, and the kind composition of
semantic properties to constructs is simply realized by textual
concatenation.

Here is an example of such use, taken directly from one of our projects
that uses semantic properties~\cite{KiniryIDebug98}.

\scriptsize
\begin{verbatim}
/**
 * Returns a boolean indicating whether any debugging 
 * facilities are turned off for a particular thread.
 *
 * @concurrency GUARDED
 * @require (thread != null) Parameters must be valid.
 * @modify QUERY
 * @param thread we are checking the debugging condition 
 * of this thread.
 * @return a boolean indicating whether any debugging 
 * facilities are turned off for the specified thread.
 * @review kiniry Are the isOff() methods necessary at all?
 **/
     
 public synchronized boolean isOff(Thread thread)
 {
   return (!isOn(thread));
 }
\end{verbatim}
\normalsize

Existing tools already use these properties for translating specifications,
primarily in the form of contracts, into run-time test code.  Reto Kramer's
iContract~\cite{Kramer98}, the University of Oldenburg's Semantic Group's
Jass tool, Findler and Felliason's contract soundness checking
tool~\cite{FindlerFelleisen01}, and Kiniry and Cheong's
JPP~\cite{KiniryCheong98} are three such tools.

\subsubsection{Eiffel}
\label{sec:eiffel}

In Eiffel, as mentioned earlier, we use indexing clauses as well as
regularly structured comments to denote semantic properties.  Using
comments as well as indexing clauses is necessary because the syntax of
Eiffel dictates that indexing clauses only appear at the top of a source
file.  The syntax of comments that use semantic properties is identical to
that of indexing clauses, thus the same parser code can be used in both
instances.  An example of such use is as follows, directly from one of our
Eiffel-based projects that uses semantic properties~\cite{EBON01}.

\scriptsize
\begin{verbatim}
indexing
   description: "The Extended BON scanner."
   project:     "The Extended BON Tool Suite"
   author:      "Joseph R. Kiniry <kiniry@acm.org>"
   copyright:   "Copyright (C) 2001 Joseph R. Kiniry"
   version:     "$Revision: 1.10 $"
   license:     "Eiffel Forum Freeware License v1"
\end{verbatim}
\normalsize

\subsection{Specification Languages}
\label{sec:spec-lang}

We have also used semantic properties to extend the BON specification
language.

\subsubsection{BON}
\label{sec:bon}

BON stands for the \emph{Business Object Notation}.  BON is described in
whole in Walden and Nerson's \emph{Seamless Object-Oriented Software
  Architecture}~\cite{WaldenNerson94}, extended from an earlier paper by
Nerson~\cite{Nerson92}.

\paragraph{Primary Aspects}
BON is an unusual specification language in that it is \emph{seamless},
\emph{reversible}, and focuses on \emph{contracting}.  BON also has both a
textual and a graphical form.

BON is \emph{seamless} because it is designed to be used during all phases
of program development.  Multiple refinement levels (high-level design with
charts, detailed design with types, and dynamism with scenarios and
events), coupled with explicit refinement relationships between those
levels, means that BON can be used all the way from domain analysis to unit
and system testing and code maintenance.

\emph{Reversibility} summarizes the weak but invertible nature of BON's
semantics.  By virtue of its design, every construct described in BON is
fully realizable in program code.  One can specify system structure, types,
contracts, events, scenarios, and more.  Each of these constructs can not
only be interpreted into program code, but program code can be interpreted
into BON.  As far as we are aware, this makes BON unique insofar as, with
proper tool support, a system specification need not become out-of-date if
it is written in BON.

Finally, BON focuses on software \emph{contracts} as a primary means of
expressing system semantics.  These contracts have exactly the same
semantics as discussed earlier with regards to object-oriented models,
because BON is an object-oriented specification language.

BON's semantics were originally specified informally using
Eiffel~\cite{Meyer88, WaldenNerson94}.  Paige and Ostroff recently provided
an analysis of BON with an eye toward a refinement-centric formal
semantics~\cite{PaigeOstroff01a, PaigeOstroff01b}.

\paragraph{BON Technologies}
BON has been, and is being, used within several commercial and Open Source
tools: Interactive Software Engineering's EiffelCase and
\myhref{http://www.eiffel.com/products/studio51/}{EiffelStudio} tools;
Ehrke's BonBon CASE tool; Steve Thompson and Roy Phillips's BONBAZ/Envision
project; Kaminskaya's BON static diagram tool~\cite{Kaminskaya01}; Paige,
Lancaric, and Ostroff's BON CASE tool; and Kiniry's EBON tool
suite~\cite{EBON01}.

The last three are particularly exciting projects because they are
currently active and have wide applicability.  Kaminskaya's and Lancaric's
tools generate textual BON, JML, Eiffel, and Java source code from a
graphical BON specification.

\subsubsection{Extended BON}
\label{sec:extended-bon}

Kiniry's EBON tool suite has a different aim.  Its secondary purpose is
similar to other previously mentioned tools, namely the generation of
documentation from BON specifications.  But its primary use is \emph{design
  model checking} for Eiffel and Java code.

BON is extended with our set of semantic properties by (a) extending the
BON language (adding new keywords and expressions), (b) using structured
comments, and (c) using indexing clauses like those in Eiffel.  More
information on these specific extensions is available at the EBON web
site~\cite{EBON01}.

\paragraph{Domain-Specific Semantics}
Translations from BON to a source language and vice-versa are to be
represented by kind theory interpretations.  This means that changes to
either side of the translation can not only be translated, but can be
checked for validity according to its (dual) model.  This
specification-code conformance (validity) checking is what we call
\emph{design model checking}.  We use this terminology because the
specification is the theory and the program code is the model, when viewed
from the logical perspective.

A change in the source code that is part of a EBON interpretation image
will automatically trigger a corresponding change in the EBON
specification.  Likewise, any change in the EBON specification will
automatically trigger a corresponding change in the source code.

Some of these translations entail more than just a transfer of information
from a specification to a comment in the source code.  For example, an
\texttt{invariant} semantic tag is interpreted not only as documentation,
but also as run-time test code.  We do not have the space in this paper to
detail this interpretation.  It follows the same lines as related tools
that support contract-based assertions in Java and Eiffel mentioned
elsewhere in this paper.

\paragraph{Belief-Driven Development}
Some BON extensions are \emph{non-reversible} because they represent system
aspects that are either very difficult or impossible to derive.  For
example, the \texttt{time\--complexity} semantic property specifies the
computational complexity of a feature.  It is (rarely) possible to extract
such information from an algorithm with automated tools.  But the fact is,
the algorithm author often knows her algorithm's complexity.  Thus, stating
the complexity as part of the algorithm specification with a semantic
property is easy and straightforward task.

Now the question arises: How do we know such specifications that are not
automatically checkable remain valid?  This is where the earlier-mentioned
belief truth structures of kind theory come into play.  

When the programmer writes the original \texttt{time\--complexity} semantic
property for a feature, she is stating a \emph{belief} about that feature.
Beliefs in kind theory are autoepistemic (the representation of the
programmer is part of the logical sentence encoding the belief), have an
associated ``strength'' or ``surety'' metric (recall
Section~\ref{sec:brief-overview-kind}), and include a set of
\emph{evidence} supporting the belief.  

We use a number of techniques to ensure that old or out-of-date beliefs are
rechecked.  With regards to this example, we define a \emph{continued
  validity condition} as part of the evidence, which is machine checkable.
Currently, if the program code or documentation to which the complexity
metric belief is bound radically changes in size, or if the feature has a
change in type, author, or other potentially complexity-impacting
specification (e.g., concurrency, space-complexity, etc.), then the
validity condition is tripped and the developer is challenged to re-check
and validate the belief, restarting the process.

\section{Experiences}
\label{sec:experiences}

We have used semantic properties within our research group, in the
classroom, and in two corporate settings.

The Compositional Computing Group at Caltech has used semantic properties
in our complex, distributed and concurrent architectures, written in Java
and Eiffel, over the past five years.  We have witnessed their utility by
first-hand experience primarily during the introduction of our complex
technologies to new students and collaborators is particularly facilitated
by semantic properties.

Students grumble at first when they are told that their comments now have a
precise syntax and a semantics.  The students initially think of this as
being ``just more work'' on their part---yet another reason to hand in a
late assignment.  But, as the term goes by, the students incorporate the
precise documentation with semantic properties and related tools into their
development process.  After a few weeks of indoctrination, they not only
stop complaining, but start praising the process and tools.  We generally
see higher quality systems and the students report spending \emph{less}
time on their homework than when they started the course.  They have
learned how to work \emph{smarter}, not \emph{harder}.

These languages, process, and tools were also used in a corporate setting
to develop a enormously complex, distributed, concurrent architecture.
When showing the system to potential funders and collaborators, being able
to present the system architecture and code with this level of
specification and documentation invariably increased our value proposition.
Uniformly, investors were not only shocked that a startup would actually
\emph{design} their system, but to think that we used lightweight formal
methods to design, build, test, and document the system was absolutely
unheard of.

We have incorporated feedback from these three domains into our work.  Our
set of semantic properties is still evolving, albeit at a rapidly
decreasing rate.  Our tools see refinement for incorporation into new
development processes, better error handling, and more complete and correct
functionality.  This user feedback is essential to understanding how these
technologies and theory can be exposed in academic and industrial settings.

\section{Conclusion}
\label{sec:conclusion}

Documentation reuse is most often discussed in the literate
programming~\cite{ChildsSametinger96} and hypertext
domains~\cite{FischerMcCallMorch89}.  Little research exists for
formalizing the semantics of semi-structured documentation.  Some work in
formal concept analysis and related formalisms~\cite{SimosAnthony98,
  Wille92} has started down this path, but with extremely loose semantics
and little to no tool support.

Recent work by Wendorff~\cite{Wendorff99, Wendorff01} bears resemblance to
this work both in its nature (that of concept formation and resolution) and
theoretic infrastructure (that of category theory).  Our work is
differentiated by its broader scope, its more expressive formalism, and its
realization in tools.  Additionally, the user-centric nature of kind theory
(not discussed in this article) makes for exposing the formalism to the
typical software engineer a straightforward practice.

Our next steps are on two fronts.  First, we are interested in embedding
our semantic properties in the Java-centric specification language JML.
Second, we are continuing to develop new tools and technologies to realize
knowledgeable development environments that use kind theory as a formal
foundation.

\subsection{JML}
\label{sec:jml}

JML is the Java Modeling Language~\cite{LeavensBakerRuby99}.  JML is a
Java- and model-centric language in the same tradition as Larch and VDM.
JML is used to specify detailed semantic aspects of Java code and some tool
support exists for type-checking and translating these specifications into
documentation and run-time test code~\cite{Bhorkar00, Ganapathy99,
  Raghavan00}.  The formal semantics of JML have been partially specified
via a logic as part of the LOOP project~\cite{JacobsPoll00}.

Extending JML with semantic properties would follow the same course that we
have used for BON.  Because we already have integrated semantic properties
with Java, and given the existing tool support for JML, we should be able
to realize inter-domain interpretations that preserve a vast amount of
information about JML-specified Java systems using kind theory.

\subsection{Social Implications}
\label{sec:social-implications}

We expect that knowledgeable development environments will have social
implications for software development.

First, this challenging, interactive style imposed by knowledgeable
development environments is not typical---we have to make sure that we are
not introducing some kind of formal methods ``paper clip''.  Thus, the
environment needs to ``tune'' itself to the interactive style and
development process of the user.  We look forward to theoretically
representing such styles in kind theory so that tuning is simply part of
the logical context.

Second, in our extensive experience in the research lab, classroom, and
corporate office, we have witnessed the fact that most developers are very
uncomfortable \emph{starting from scratch}, especially with regards to
system documentation and informal and formal specifications.  If some
existing documentation or specification exists, developers are much more
likely to continue in that trend because they feel that they are
\emph{contributing} rather than \emph{creating}.  

Because the EBON tool suite will automatically generate a base
specification from program code, and because the specification-code
validity conformance is automatically maintained, we have a primer as well
as a positive feedback cycle for lightweight specification with semantic
properties.  Only time and experience will tell whether this is a
sufficient fire to light the correct software fuse.

\subsection{Knowledgeable Environments}
\label{sec:knowl-envir}

As mentioned previously, our work on KDEs continues.  

We wish to augment our already powerful development environment in two
ways.  Our first step entails integrating an interactive front-end like
XEmacs with our kind system realized in Maude.  The availability of
Emacs-centric tools for proof system like the excellent
\myhref{http://zermelo.dcs.ed.ac.uk/$\sim$proofgen/}{Proof General} make
this a relatively straightforward exercise.  The most time-consuming aspect
is writing interpretation engines that translate annotated source code to
and from a kind representation format.  Several such tools are being
prototyped now~\cite{EBON01, KiniryCheong98}.

We also plan on integrating these environments with our reusable knowledge
repository known as the Jiki~\cite{Jiki98}.  The Jiki is a read/write web
architecture realized as a distributed component-based Java Wiki.  All
documents stored in the Jiki are represented as instances of kind.
Manipulating Jiki assets, including adding or deleting information or
searching for reusable assets, is realized through a forms-based web
interface as well as through a Java component-based API.

\section{Acknowledgments}

This work was supported under ONR grant JJH1.MURI-1-CORNELL.MURI (via
Cornell University) ``Digital Libraries: Building Interactive Digital
Libraries of Formal Algorithmic Knowledge'' and AFOSR grant
JCD.\-61404-1-AFOSR.\-614040 ``High-Confidence Reconfigurable Distributed
Control''.


\bibliographystyle{plain}



\appendix

\section{Semantic Properties Summary}

\begin{table}[htbp]
  \caption{The Full Set of Semantic Properties}
  \label{tab:The_Full_Set_of_Semantic_Properties}
  \begin{center}
    \begin{tabular}{|ccc|}
      \hline
      \textbf{Meta-Information:} & \textbf{Contracts}    & \textbf{Versioning} \\
      author                     & ensure                & version \\
      bon                        & generate              & deprecated \\
      bug                        & invariant             & since \\
      copyright                  & modifies              & \textbf{Documentation} \\
      description                & require               & design \\
      history                    & \textbf{Concurrency}  & equivalent \\
      license                    & concurrency           & example \\
      title                      & \textbf{Usage}        & see \\
      \textbf{Dependencies}      & param                 & \textbf{Miscellaneous} \\
      references                 & return                & guard \\
      use                        & exception             & values \\
      \textbf{Inheritance}       & \textbf{Pending Work} & time-complexity \\
      hides                      & idea                  & space-complexity \\
      overrides                  & review                &  \\
                                 & todo                  &  \\
      \hline
    \end{tabular}
  \end{center}
\end{table}


\end{document}